\documentclass[twocolumn]{aastex63}

\usepackage{ulem}
\usepackage{amsmath}

\shorttitle{JAGB Gaia ED3 Calibration}
\shortauthors{Lee et al.}
\graphicspath{{./}{figures/}}

\begin{document}

\title{\textbf{A Preliminary Calibration of the JAGB Method Using \textit{Gaia} EDR3}}

\author{Abigail~J.~Lee}\affil{Department of Astronomy \& Astrophysics, University of Chicago, 5640 South Ellis Avenue, Chicago, IL 60637}\affiliation{Kavli Institute for Cosmological Physics, University of Chicago,  5640 South Ellis Avenue, Chicago, IL 60637}

\author{Wendy~L.~Freedman}\affil{Department of Astronomy \& Astrophysics, University of Chicago, 5640 South Ellis Avenue, Chicago, IL 60637}\affiliation{Kavli Institute for Cosmological Physics, University of Chicago,  5640 South Ellis Avenue, Chicago, IL 60637}

\author{Barry~F.~Madore}\affil{Observatories of the Carnegie Institution for Science 813 Santa Barbara Street, Pasadena, CA~91101}\affil{Department of Astronomy \& Astrophysics, University of Chicago, 5640 South Ellis Avenue, Chicago, IL 60637}

\author{Kayla~A.~Owens}\affil{Department of Astronomy \& Astrophysics, University of Chicago, 5640 South Ellis Avenue, Chicago, IL 60637}

\author{In Sung Jang}\affil{Department of Astronomy \& Astrophysics, University of Chicago, 5640 South Ellis Avenue, Chicago, IL 60637}\affiliation{Kavli Institute for Cosmological Physics, University of Chicago,  5640 South Ellis Avenue, Chicago, IL 60637}

\correspondingauthor{Abigail J. Lee}
\email{abbyl@uchicago.edu}

\begin{abstract}
The recently-developed J-region Asymptotic Giant Branch (JAGB) method has extraordinary potential as an extragalactic standard candle, capable of calibrating the absolute magnitudes of locally-accessible Type Ia supernovae, thereby leading to an independent determination of the Hubble constant.
Using \textit{Gaia} Early Data Release 3 (EDR3) parallaxes, we calibrate the zeropoint of the JAGB method, based on the mean luminosity of a color-selected subset of carbon-rich AGB stars. 
We identify Galactic carbon stars from the literature and use their near-infrared photometry and \textit{Gaia} EDR3 parallaxes to measure their absolute $J$-band magnitudes. Based on these Milky Way parallaxes we determine the zeropoint of the JAGB method to be $M_J=-6.14 \pm0.05$ (stat) $\pm$ 0.11 (sys)~mag. 
This Galactic calibration serves as a consistency check on the JAGB zeropoint, agreeing well with previously-published, independent JAGB calibrations based on geometric, Detached-Eclipsing Binary (DEB) distances to the LMC and SMC.
However, the JAGB stars used in this study suffer from the high parallax uncertainties that afflict the bright and red stars in EDR3, so we are not able to attain the higher precision of previous calibrations, and ultimately will rely on future improved DR4 and DR5 releases.
\end{abstract}


\section{Introduction}
Measuring the Hubble constant ($H_0$) precisely and accurately has remained a challenging endeavor for observational cosmologists. 
$H_0$  parameterizes the current expansion rate of the universe, and together with measurements of the cosmic microwave background radiation, it provides a powerful constraint on the standard model of cosmology \citep[e.g.,][]{2010ARA&A..48..673F, 2020A&A...641A...6P}.
Because of its significance, the value of $H_0$ has undergone intense scrutiny and re-determination over the past few decades.
Currently, early universe measurements of $H_0$ (forward modeled to the present day) as inferred from the measurements of the power spectra of temperature and polarization anisotropies of the Cosmic Microwave Background (CMB) appear to differ by $4-6 \sigma$ from measurements directly measured by late-time universe distance ladders \citep{2019NatAs...3..891V}. Even more perplexingly, the two most accurate and precise late-time universe values of $H_0$ measured by the SHoES Group via the Leavitt Law \citep{2016ApJ...826...56R, 2019ApJ...876...85R, 2021ApJ...908L...6R} and by the Carnegie-Chicago Hubble Program (CCHP) via the Tip of the Red Giant Branch (TRGB) method \citep{2019ApJ...882...34F, 2020ApJ...891...57F, 2021arXiv210615656F} differ by about 2-$\sigma$. Clearly, the local universe tension should be reconciled before concluding unambiguously that new physics is required to bridge the gap between early- and late-universe measurements of $H_0$. 

Recently, a new independent distance indicator was proposed as an alternative Type Ia supernovae calibrator to Cepheids and the TRGB in order to act as a `tie-breaker,' especially for more distant galaxies where the agreement between the two standard methods begins to break down, and where hidden systematic uncertainties may be revealed. This new method, the \textit{J-region} Asymptotic Giant Branch (JAGB) method, was introduced by \cite{2020ApJ...899...66M}, further applied and tested by \cite{2020ApJ...899...67F}, and also independently developed by \cite{2020MNRAS.495.2858R}. 

JAGB stars are thermally-pulsating asymptotic giant branch (TP-AGB) stars that are undergoing dredge-up events: instabilities in the two energy-producing shells in these stars induce episodes of upper-envelope convection that eventually penetrate deep into the helium-fusing shell that can result in freshly-produced carbon being brought to the surface. The carbon then gives these stars a much redder appearance than their bluer oxygen-rich counterparts \citep{2003agbs.conf.....H}. 
In order to distinguish this region of C-rich stars from the O-rich AGB stars to the blue and the extreme carbon stars to the red, \cite {2000ApJ...542..804N} defined color limits of $1.40<(J-K_s)<2.00$~mag for this class of stars they labeled as `Region J' in a $K$ versus $(J-K_s)$  color-magnitude diagram. The JAGB stars are also bounded in luminosity: JAGB stars have an \textit{upper} magnitude limit as more luminous (and thus more massive) stars undergo ``Hot-Bottom Burning," where as the carbon is being transported to the surface of the star during dredge-ups, the star is so hot the carbon is converted to nitrogen before it can reach the atmosphere. There also exists a \textit{lower} magnitude limit because the dredge-up events for fainter (and thus less massive) stars are ineffective in transporting carbon to the surface. Therefore, JAGB stars have well defined color and magnitude limits and can be easily photometrically identified.

Over 15 years ago, the mean absolute I-band magnitude of carbon stars was shown to be remarkably stable from galaxy to galaxy and was therefore proposed as a standard candle by \citet{2005A&A...434..657B}.
Recently,  \cite{2020arXiv201204536L} compared distances determined by the JAGB method, Cepheid Leavitt Law, and TRGB in the local group galaxy Wolf-Lundmark-Melotte, finding agreement at the 3\% level amongst the three methods.  \cite{2020arXiv201204536L} also found that the JAGB method had comparable precision to the Cepheid Leavitt law and TRGB, indicating enormous potential for its own future, independent $H_0$ measurement. 

\cite{2020ApJ...899...66M} were the first to calibrate the JAGB method in the LMC and SMC via Detached-Eclipsing Binary (DEB) distances \citep{2019Natur.567..200P, 2014ApJ...780...59G}. In this paper, we provide an independent consistency check on that calibration using the \textit{Gaia} EDR3 parallaxes.
\textit{Gaia} is a European Space Agency satellite that recently provided accurate astrometry for over 1.8 billion nearby sources in their Early Data Release 3 (EDR3) \citep{2016A&A...595A...1G, 2020arXiv201201533G}. 
However, limits in the accuracy of the \textit{Gaia} parallaxes are still present, especially for bright and red sources, which we discuss in Section \ref{sec:sec4}. Hence, our \textit{Gaia} calibration remains preliminary only, and we expect that it will be superseded by future \textit{Gaia} releases.

The outline of this paper is as follows: in Section \ref{sec:sec2}, we discuss the photometry of our sample of Galactic JAGB stars. In Section \ref{sec:sec3}, we give an overview of the \textit{Gaia} distance measurements used in this study. In Section \ref{sec:sec4}, we present an absolute calibration of the JAGB Method zeropoint via \textit{Gaia} parallaxes, and discuss the limitations of the \textit{Gaia} parallaxes for bright and red stars. In Section \ref{sec:sec5}, we compare our absolute calibration with previous calibrations. Finally, in Section \ref{sec:summary}, we present a summary of this paper and the future direction of the JAGB method.

\section{Photometry}\label{sec:sec2}
In this section, we describe the Milky Way carbon stars used in this study. We identified two robust catalogs of Galactic carbon stars from \cite{2006MNRAS.369..751W} and \cite{2012AJ....143...36C}.

In Appendix \ref{App:fail}, we also describe our attempts to identify JAGB stars solely on the basis of their color from the 2MASS Point-Source Catalog \citep{2006AJ....131.1163S}. 
Unfortunately, stars in the 2MASS catalog with $m_J<5.5$~mag, a magnitude limit that many Galactic JAGB stars are brighter than,
have large photometric uncertainties on the order of $0.25$~mag \citep{2006AJ....131.1163S}. \textit{Gaia} also reported potential large systematic parallax uncertainties for both bright and red stars, of which JAGB stars are both \citep{2020arXiv201203380L}.
In conclusion, we found the large photometric and parallax uncertainties of bright JAGB stars made it difficult to discern them solely on the basis of color.

\subsection{The Milky Way Carbon Star Sample}\label{subsec:catalog}

\cite{2006MNRAS.369..751W} identified 239\footnote{We chose to exclude the 18 stars in their list which were labeled as peculiar or unlikely bona fide carbon stars.} Galactic carbon-rich variable stars from the IRAS Point Source Catalogue \citep{1988iras....1.....B}, the \cite{1989ApJS...70..637A} carbon star catalog, and their own observations of C stars, for which they measured $JHKL$ photometry, defined on the South African Astronomical Observatory (SAAO) magnitude system \citep{1990MNRAS.242....1C}, which we converted to the 2MASS magnitude system using transformations from \cite{2007MNRAS.380.1433K}.
These carbon stars were selected on the basis of their color, $K$-band variability, carbon star SiC emission feature at 11.2 $\mu$m, 25-to-12 $\mu$m flux ratio, and 12 $\mu$m flux.
The goal of their study was to identify Galactic carbon stars with accurate distances and radial velocities in order to gain insight into the ages and masses of the local carbon star population.\footnote{We removed 32 stars from the \cite{2006MNRAS.369..751W} catalog that were missing $J$-band photometry.}

\cite{2012AJ....143...36C} compiled a sample of 972\footnote{There were 2 objects, IRAS 03291+4116 and IRAS 20204+2914, for which there were duplicate entries. However, neither star resided in the JAGB star color range so they ended up being inconsequential to the final calculation.} Galactic carbon stars from the literature, with 2MASS $JHK_s$ photometry. These stars were selected by their color, carbon star SiC emission feature at 11.2 $\mu$m, flux ratios, and HCN + $\textrm{C}_2\textrm{H}_2$ feature in absorption spectra at 3.1 $\mu$m. The goal of this study was to investigate carbon-rich AGB stars in the late stages of evolution.

Both \cite{2012AJ....143...36C} and  \cite{2006MNRAS.369..751W} applied Galactic interstellar extinction corrections to their photometry, using the \cite{2003A&A...409..205D} three-dimensional Galactic extinction model, and following the reddening law from \cite{1999hia..book.....G}.

\section{Gaia EDR3 Distances}\label{sec:sec3}

To obtain distance estimates for each source, we chose not to perform a simple cone search between our carbon star catalog and the \textit{Gaia} EDR3 catalog, as generally the most probable match is often not necessarily the nearest neighbor. 
Instead, the \textit{Gaia} team has already matched the \textit{Gaia} EDR3 catalog with the 2MASS catalog, calculating the most probable neighbor based on positional and source density properties \citep{2017A&A...607A.105M, 2019A&A...621A.144M}. This catalog is publicly available\footnote{Find the documentation at \url{https://gea.esac.esa.int/archive/documentation/GEDR3/Gaia_archive/chap_datamodel/sec_dm_crossmatches/ssec_dm_tmass_psc_xsc_best_neighbour.html}} and can be queried at \url{https://gaia.aip.de/query/}.
To find each carbon star's corresponding 2MASS identifier, we located its \textsc{Simbad} entry using its R.A./Dec. coordinates and name. Every star in the  \cite{2006MNRAS.369..751W} catalog had a corresponding 2MASS identifier. 14/972 stars in the \cite{2012AJ....143...36C} catalog lacked 2MASS identifiers, and were removed.

Using the \texttt{tmass\_psc\_xsc\_best\_neighbour} catalog, we then determined the unique \textit{Gaia} EDR3 \texttt{source id} for each star from its corresponding 2MASS identifier. There were several sources in both catalogs that lacked \textit{Gaia} counterparts: two in the \cite{2006MNRAS.369..751W} catalog and 85 in the \cite{2012AJ....143...36C} catalog. This may occur because as carbon stars are moderately dusty, they may be too faint in the optical for \textit{Gaia} to detect. This is one reason why it was beneficial to use this procedure instead of a simple cone search; while a cone search will always result in a nearest neighbor, utilizing the \textit{Gaia} team's best-neighbor match catalog revealed whether a given source had a genuine Gaia counterpart.

We then combined the two catalogs, finding 108 sources in common between them. This brought the total number of unique sources in our list to 970. If a source overlapped in both lists, we averaged its flux.

Then, using each star's \textit{Gaia} EDR3 {\tt source id}, we acquired a distance to each star from the  \cite{2020arXiv201205220B} catalog\footnote{Publicly available at \url{https://dc.zah.uni-heidelberg.de/gedr3dist/q/cone/form}} of stellar distances, which was derived from \textit{Gaia} EDR3 information. 30/970 stars did not have \cite{2020arXiv201205220B} distances available, and were removed.
The catalog gives two types of distance estimates for a given source: the \textit{geometric} and \textit{photogeometric} distance. The geometric distances were calculated from the star's parallax, parallax uncertainty, and Galactic latitude and longitude. The photogeometric distances were calculated from the aforementioned parameters, as well as the star's apparent $G$-band magnitude and $BP-RP$ color.
\cite{2020arXiv201205220B} noted their model for calculating photogeometric distances failed for stars at the bluest and reddest ends of the color range [ ($BP-RP$) $\lesssim0$ or ($BP-RP$) $\gtrsim4$~mag, respectively; see Figures 4 and 5 of their paper]. As carbon stars are extremely red stars, we found a fraction (7\%) of our carbon star sample had ($BP-RP>4$)~mag, and thus lacked photogeometric distances. Therefore, we opted to use the geometric distances.

A common way of selecting sources with reliable astrometry from \textit{Gaia} is to only use parallaxes much larger than their estimated uncertainties \citep{2020arXiv201206242F}. We eliminated sources with $\sigma_{\pi}/\pi>0.20$, where $\pi$ is the measured parallax by \textit{Gaia} and $\sigma_{\pi}$ is the uncertainty on the parallax. 
The \textit{Gaia} team also recommends cutting on the parameter {\tt ruwe}, the primary indicator for the quality of a given astrometric solution \citep{2020arXiv201206242F}.
We used a cut of {\tt ruwe}$<2.0$, determined by \cite{2021arXiv210110206M} to be a likely safe cut for determining high-quality parallaxes. After making these two data cuts, 585 carbon stars remained in our final catalog. We have made this catalog publicly available and show a sample of it in Table \ref{tab:catalog}.
Finally, we calculated the absolute $J$-band magnitude $M_J$ for each star, using its apparent $J$-band magnitude and distance from the \cite{2020arXiv201205220B} catalog.

\begin{deluxetable*}{cccccc}
\tablenum{1}
\tablecaption{Carbon star catalog
\label{tab:catalog}}
\tablewidth{1pt}
\tablehead{
\colhead{\textit{Gaia} EDR3 source\_id} & 
\colhead{2MASS ID} & 
\colhead{$J$ [mag]} & 
\colhead{$(J-K_s)$ [mag]} & 
\colhead{Parallax [mas]} & 
\colhead{Distance\tablenotemark{\scriptsize a} [pc]} 
}
\startdata
5118511817421484544 & $02291531-2605559$ & 4.074 & 2.536 & 1.807 & 540.0 \\
3314149224049447040 &  $04312193+1739103$ &  6.337 &  2.715 &  0.671 &  1454.6
\\
2901319097862449536 &  $05434305-3223287$ &  9.402 &  3.573 &  0.571 &  1573.3
\\
3400017749285334784 &  $05453941+2041420$ &  2.031 &  1.705 & 1.501 &  649.7
\\
3340248194120411776 &  $05453669+1216152$ &  6.277 & 2.261 &  0.318 &  2728.1
 \\
3373737664738495744 &  $06114782+1908200$ &  8.225  & 3.342 &  0.577 &  1536.4
\\
3325351357754529152 & $06215800+0720576$ &  4.413  & 2.068 &  0.662 & 1442.4
\enddata
\tablenotetext{\scriptsize a}{Geometric distance from \cite{2020arXiv201205220B}}
\tablecomments{Table 1 is published in its entirety in the machine-readable format. A portion is shown here for guidance regarding its form and content.}
\end{deluxetable*}

\section{Absolute Calibration of the JAGB Zero point}\label{sec:sec4}

\begin{figure*}
\centering
\includegraphics[width=\textwidth]{"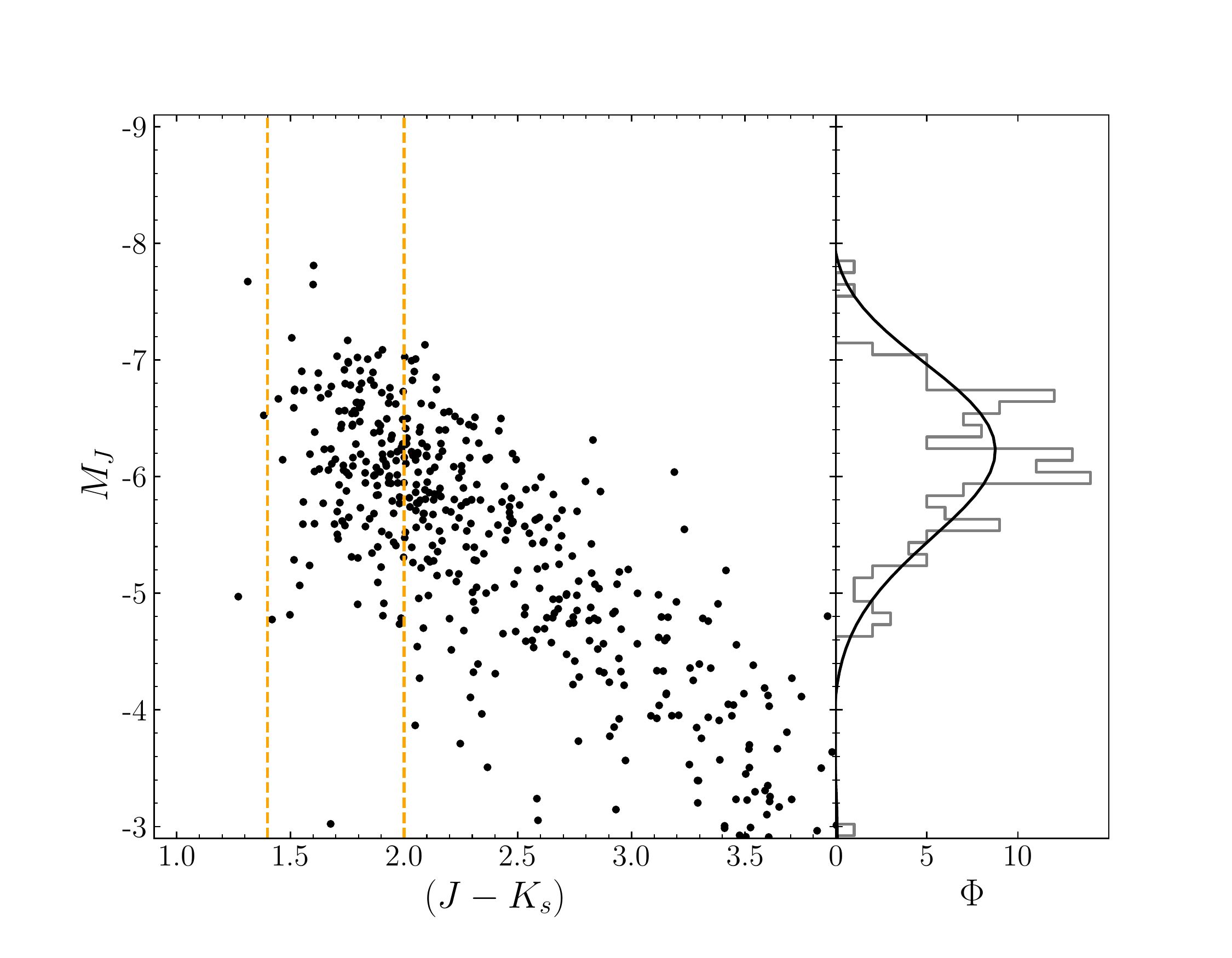"}
\caption{Color Magnitude Diagram (left panel) and GLOESS smoothed $J$-band luminosity function (right panel) for our sample of Milky Way carbon stars. The smoothed luminosity function is only shown for illustrative purposes and was not used in any calculations. The JAGB sample of stars between $1.40<(J-K_s)<2.00$~mag is clearly defined. The extreme carbon stars, which cannot be used as standard candles, are excluded with our red color limit of $(J-K_s)=2.0$~mag.}
\label{fig:f4}
\end{figure*}
We present our final $J$ vs. $(J-K_s)$ color magnitude diagram and $J$-band luminosity function for our sample of 585 carbon stars in Figure \ref{fig:f4}. While this is a carbon star sample, not all carbon stars are JAGB stars. Extreme carbon stars with $(J-K_s)>2.0$~mag do not have a constant J magnitude, but decrease in magnitude with increasing $(J-K_s)$ color. These extreme carbon stars are heavily affected by dust obscuration and strong winds \citep{2011AJ....142..103B}, and have been excluded from our JAGB star sample after making the color cuts of $1.40<(J-K_s)<2.00$~mag following \cite{2000ApJ...542..804N}. 
Based on the 153 remaining JAGB stars, we measured the median absolute $J$-band magnitude for the JAGB stars to be $M_J = -6.14$~mag. We chose not to utilize the mean as has been used in previous papers \citep{2020ApJ...899...66M, 2020ApJ...899...67F, 2020arXiv201204536L}, as there were three outliers $>-4$~mag and $<-8$~mag that significantly biased the mean $M_J$ to $-6.03$~mag (eliminating these three outliers gave a mean $M_J=-6.15$~mag). We plan to explore the merits of using the mean, the median, and other methods for determining the JAGB magnitude in a future program aimed at obtaining high-precision $JHK$ photometry for several nearby galaxies.
Finally, we discuss the statistical and systematic uncertainties associated with this measurement in Section \ref{subsec:parallaxsigma} below.

\subsection{Gaia Parallax Uncertainties for Bright and Red Stars}\label{subsec:parallaxsigma}

\cite{2001ApJ...548..712W} were the first to utilize carbon stars as a distance indicator in the broad-band near-infrared filters. For their `Region J' stars, they measured the scatter to be $\pm0.38$~mag about their $K_s$ vs. ($J-K_s$) relation, using single-epoch data. In comparison, the observed scatter on our JAGB star sample's median $M_J$ (excluding the three outliers $>-4$~mag and $<-8$~mag) was measured to be $\pm0.60$~mag, yielding an error on the mean of $0.60/\sqrt{150} = 0.05$~mag, which we adopt as a statistical uncertainty on our measurement.

We note that the scatter in the observed luminosity function must include contributions from the intrinsic variability of the JAGB stars ($\sim 0.2$~mag), the variance due to extinction, parallax uncertainties, and photometric uncertainties \citep{2001ApJ...548..712W}. The \cite{2006MNRAS.369..751W} $J$-band photometry has a precision better than $\pm0.03$~mag and the median photometric uncertainty of the \cite{2012AJ....143...36C} $J$-band photometry is $\pm0.026$~mag, indicating that the photometric uncertainties contribute little to the observed scatter.

Next, we estimated the uncertainty introduced from the extinction corrections applied to our photometry. \cite{2006MNRAS.369..751W} provided $A_V$ values determined by their models for 144 C-rich Miras. We found that 17 of these Miras were also JAGB stars and passed our data cleaning cuts (\texttt{ruwe>2.0} and $\sigma_{\pi}/\pi<0.20$), with a median extinction value in the $J$ band of $A_J=0.16$~mag.\footnote{Assuming $A_J=0.272 A_V$ \citep{1998ApJ...500..525S}.} 
The Carnegie-Chicago Hubble Program procedure is to adopt half of the reddening value as its systematic uncertainty, which in this case would then be 0.08~mag (e.g., \citealt{2020arXiv200804181J}). However, even adopting the full median extinction value of $A_J=0.16$~mag as an uncertainty would still not be enough to explain the large observed scatter in the absolute $J$-band luminosity function.

Therefore, the large scatter may be explained, in part, by the fact that the astrometric precision of the \textit{Gaia} parallaxes in EDR3  has strong color and magnitude dependencies, relying heavily on accurate PSF fitting and calibration \citep{2020arXiv201202069R}. While we found the median parallax uncertainty of our sample of JAGB stars to be only $\pm0.027$~mas, recent publications have found the \textit{Gaia} EDR3 parallax uncertainties to be underestimated for brighter stars \citep{2021MNRAS.506.2269E}.
For stars brighter than $G=13$~mag, the instrument switches its main CCD sampling scheme to mitigate saturation effects. However, this secondary mode is more sensitive to PSF modeling deficiencies, leading to larger uncertainties \citep{2020arXiv201202069R}. These uncertainties are relevant for not only the bright JAGB stars, as all of the JAGB stars used in this study were brighter than $G=13$~mag, but other local distance indicators\footnote{For example, the Cepheid sample of \cite{2021ApJ...908L...6R} recently used to calibrate the Leavitt Law using \textit{Gaia} EDR3 is brighter than $G=11.2$~mag.} as well.
Fortunately, as detailed by \cite{2020arXiv201202069R}, improvements in the PSF modeling, particularly for bright stars ($G<13$~mag), are already being implemented for DR4. 
Advancements in the PSF modeling for \textit{Gaia} DR4 will benefit the astrometry for several distance indicators including the JAGB stars and Cepheids. 

In addition to the large statistical uncertainties, the \textit{Gaia} parallaxes for the JAGB stars also likely suffer from systematic offsets. 
Although in theory, \textit{Gaia} should be able to measure absolute parallaxes: that is, for sufficiently distant sources such as quasars, the measured parallax by \textit{Gaia} should be zero. However, the accuracy of \textit{Gaia}'s astrometry has been shown to be subject to various instrumental effects which cause a global systematic offset in the parallaxes with respect to the International Celestial Reference System \citep{2018A&A...616A...2L}, referred to as the \textit{parallax zeropoint} offset.
For their Early Data Release 3 in December 2020, \textit{Gaia} reported a median zeropoint offset of $-17$~$\mu$as, with systematic variations on the order of $10$~$\mu$as, defined by over 1.6 million background quasars 
\citep{2020arXiv201203380L}.
However, this parallax zeropoint offset was derived from quasars in the magnitude and color range $G>16$~mag and $(GP-RP)<1.6$~mag \citep{2020arXiv201203380L}, respectively, a range within which none of our bright and red JAGB stars fall. \cite{2020arXiv201203380L} speculated that sources outside of the region of parameter space well populated by quasars could have additional systematic uncertainties ranging from ``a few $\mu$as'' to ``several tens of $\mu$as.'' 

Several recent publications have attempted to also quantify additional parallax offsets in the \textit{Gaia} EDR3 data. Using first-ascent red giant branch stars with asteroseismic parallaxes,
\cite{2021arXiv210107252Z} derived an offset of $-15 \pm3~\mu$as. \cite{2021arXiv210109691H} found a global offset of $-9.8 \pm1.0~\mu$as for the $<10.8$ mag LAMOST primary red clump (PRC) stellar sample. \cite{2021ApJ...907L..33S} compared the EDR3 parallaxes with distances derived from eclipsing binaries and found a maximum additional offset of $+33$ $\mu$as. Using globular clusters, \cite{2021arXiv210110206M}, measured an offset of $+6.9\pm2.2~\mu$as  for  stars  in the  range  $9.3< G<11$~mag. Finally, \cite{2021ApJ...908L...6R} derived a $-14 \pm 6~\mu$as offset by comparing the EDR3 parallaxes with HST photometrically-predicted parallaxes.
Thus, due to the lack of consensus on a parallax offset value, we conservatively adopt $\pm30$ $\mu$as as the systematic uncertainty on the parallax measurements.
We propagated this uncertainty on the parallax estimates into an error on $M_J$, which amounted to $\pm0.11$~mag. We thus adopt $\pm0.11$~mag as a systematic uncertainty on the absolute calibration. 

We look forward to the \textit{Gaia} DR4, which promises better astrometry and self-calibration of the global parallax zeropoint, especially for brighter and redder stars \citep{2020arXiv201203380L}. In conclusion, we report our final value for the JAGB zeropoint to be $M_J=-6.14 \pm0.05$ (stat) $\pm$ $0.11$ (sys) mag. 

\section{Comparison with Previous Calibrations}\label{sec:sec5}

\begin{deluxetable*}{ccc}
\tablenum{2}
\tablecaption{Previous JAGB Calibrations
\label{tab:compare}}
\tablewidth{1pt}
\tablehead{
\colhead{Study} & 
\colhead{$M_J$ [mag]} & 
\colhead{Anchor}
}
\startdata
\cite{2020ApJ...899...66M} & $-6.22\pm0.01$ (stat) $\pm$ 0.03 (sys) &  LMC DEB \citep{2019Natur.567..200P}\\
\cite{2020ApJ...899...66M} & $-6.18\pm0.01$ (stat) $\pm$ 0.05 (sys) & SMC DEB \citep{2014ApJ...780...59G}\\ 
\cite{2020MNRAS.495.2858R} & $-6.28\pm0.004$ & LMC DEB \citep{2019Natur.567..200P}\\
\cite{2020MNRAS.495.2858R} & $-6.16\pm0.015$ & SMC Leavitt Law \citep{2016ApJ...816...49S}\\
\cite{2020MNRAS.495.2858R} & $-5.60\pm0.026$ & \textit{Gaia} DR2 Galactic Parallaxes\\
This Work & $-6.14 \pm0.05$ (stat) $\pm $ 0.11 (sys) & \textit{Gaia} EDR3 Galactic Parallaxes
\enddata
\end{deluxetable*}

Previous JAGB calibrations are compiled in Table \ref{tab:compare}. This is the first study to date that has utilized the \textit{Gaia} EDR3 parallaxes as the geometric anchor for the JAGB method. Two previous studies, \cite{2020ApJ...899...66M} and \cite{2020MNRAS.495.2858R} both anchored the JAGB method to the LMC and SMC.  \cite{2020MNRAS.495.2858R} also measured a JAGB zeropoint based on \textit{Gaia} DR2 parallaxes.

For the LMC-based calibrations, \cite{2020ApJ...899...66M} found $M_J = -6.22$ $\pm$ 0.01 (stat) $\pm$ 0.03 (sys) mag, based on a DEB distance to the LMC \citep{2019Natur.567..200P}. Using the same LMC DEB distance, \cite{2020MNRAS.495.2858R} measured $M_J=-6.28\pm0.004$~mag. 

For the SMC-based calibrations, \cite{2020ApJ...899...66M} found $-6.18\pm0.01$ (stat) $\pm$ 0.05 (sys) mag based on the SMC DEB distance from \cite{2014ApJ...780...59G}. \cite{2020MNRAS.495.2858R} found $-6.16$  $\pm$ $0.015$~mag based on the \cite{2016ApJ...816...49S} Leavitt Law SMC distance.

Finally, \cite{2020MNRAS.495.2858R} measured an $M_J=-5.60\pm0.026$~mag based on Galactic carbon stars. This calibration was measured using \textit{Gaia} DR2 parallaxes. They attributed their considerably fainter calibration to the high parallax uncertainties from \textit{Gaia} DR2. 

\section{Summary and Conclusions}\label{sec:summary}
The JAGB method is a geometrically calibrated and  simply-applied stellar distance indicator that capitalizes on the fact that JAGB stars are easily identified in the near infrared. 
We have identified a sample of 153 JAGB stars from Milky Way carbon star catalogs, for which we derived distances using \textit{Gaia} EDR3 parallaxes. We present an absolute calibration of the JAGB method, $M_J=-6.14 \pm0.05$ (stat) $\pm$ 0.11 (sys)~mag for this Galactic sample. 

Our Milky Way JAGB calibration is completely independent of, and agrees well with the \cite{2020ApJ...899...66M} LMC and SMC DEB-based calibrations of $M_J (LMC) = -6.22 \pm 0.01$ (stat) $\pm$ 0.03 (sys)~mag and $M_J (SMC) =-6.18 \pm 0.01$ (stat) $\pm$ 0.05 (sys)~mag. Averaging these three calibrations yields an updated JAGB zero-point of $M_J=-6.18 \pm0.02$ (stat) $\pm$ 0.04 (sys)~mag. Moreover, no statistically significant trend with increasing host-galaxy metallicity is found here, with the SMC ($M_J = -6.18$~mag, [Fe/H]$\sim-0.8$ dex), LMC ($M_J = -6.22$~mag, [Fe/H]$\sim-0.5$ dex), and Milky Way ($M_J = -6.14$~mag, [Fe/H]$\sim 0$ dex) already spanning a large range in present-day metallicity. 
We also note that our calibration is also in excellent agreement with the independent \cite{2020MNRAS.495.2858R} SMC calibration of $-6.16\pm0.015$~mag, and consistent to within $1.2 \sigma$ of their LMC calibration of $-6.28\pm0.004$~mag.

We have demonstrated that the JAGB method's absolute J-band calibration is consistent across three independent zero-point calibrators.  With this increasing accuracy, the stage is being set for the JAGB method to be more widely applied as a primary  extragalactic distance indicator. We look forward to the release of \textit{Gaia} DR4 in 2022, which will bring with it several improvements in its astrometry. 

\acknowledgments

A. J. L. would like to thank Andrew Madigan for providing helpful comments on code used in this work. 
We thank Rachel Wagner-Kaiser for making her Hess Diagram Python code publicly available on Github.
We thank the {\it Observatories of the Carnegie Institution for
Science} and the {\it University of Chicago} for their support of our long-term research into the calibration and determination of the expansion rate of the Universe. 
Finally, we thank the anonymous referee for their constructive suggestions.

This work has made use of data from the European Space Agency (ESA) mission
{\it Gaia} (\url{https://www.cosmos.esa.int/gaia}), processed by the {\it Gaia}
Data Processing and Analysis Consortium (DPAC,
\url{https://www.cosmos.esa.int/web/gaia/dpac/consortium}). Funding for the DPAC
has been provided by national institutions, in particular the institutions
participating in the {\it Gaia} Multilateral Agreement. 

This research made use of the cross-match service and SIMBAD database \citep{2000A&AS..143....9W}, provided by and operated at CDS, Strasbourg. This research has made use of NASA's Astrophysics Data System Bibliographic Services. 

\software{Astropy \citep{2013A&A...558A..33A, 2018AJ....156..123A}, TOPCAT \citep{2005ASPC..347...29T}, NumPy \citep{2011CSE....13b..22V}, Matplotlib \citep{2007CSE.....9...90H}, Pandas \citep{pandas}, 
HessDiagram (\url{https://github.com/rwk506/HessDiagram/})}

\facility{Gaia}

\appendix
\section{Identifying JAGB stars with 2MASS photometry}\label{App:fail}

In addition to identifying JAGB stars from spectroscopically selected carbon star catalogs in the literature, we also attempted to photometrically select JAGB stars solely on the basis of their color.
This approach follows that of \cite{2020ApJ...899...66M}, who photometrically selected JAGB stars without using carbon star catalogs for their LMC and SMC DEB-based calibrations.  

\subsection{2MASS Sample}
To build a Galactic near-infrared color magnitude diagram, $JHK_s$ data for Galactic stars were obtained from the 2MASS All-Sky Point Source Catalog \texttt{X} Gaia EDR3 best neighbor cross-match catalog (\texttt{tmass\_psc\_xsc\_best\_neighbour})
using a Galactic latitude constraint of \texttt{|l|>30}.
We chose to limit our catalog to sources with high Galactic latitude in order to avoid high line-of-sight extinction contamination from Galactic dust and source crowding \citep{2006AJ....131.1163S}. 
We also applied more stringent data cuts (\texttt{ruwe<1.4} and $\sigma_{\pi}/\pi<0.10$) than in Section \ref{sec:sec3} to obtain the highest-quality parallaxes.

\subsection{Issues in the Astrometry and Photometry}
\begin{figure*}
\figurenum{A1}
\centering
\includegraphics[width=0.5\textwidth]{"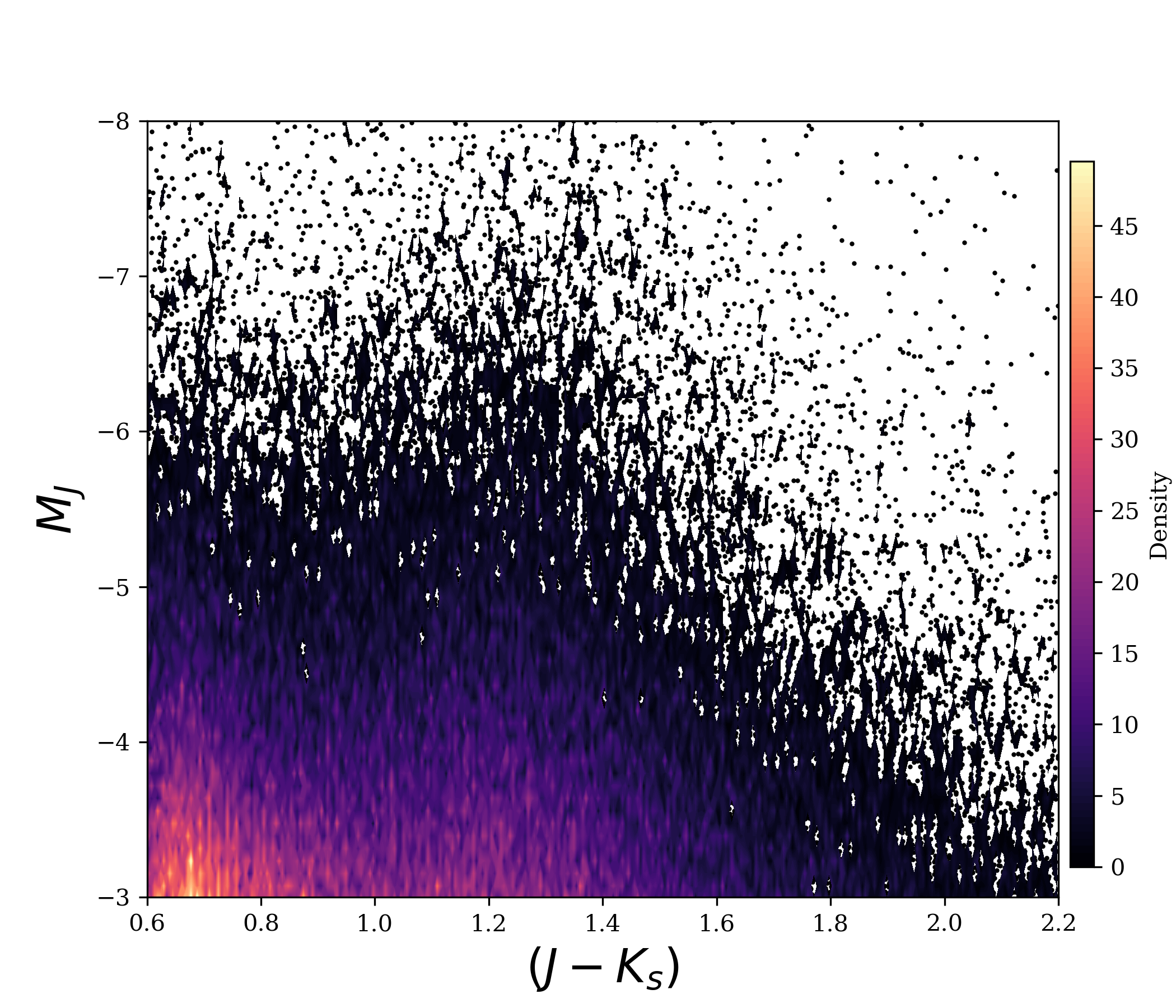"}
\caption{$J$ vs. ($J-K_s$) color magnitude Hess diagram for the full 2MASS-selected sample.  The JAGB stars are obscured by O-rich AGB stars likely because of large  errors from reddening, \textit{Gaia} parallax uncertainties, and 2MASS photometric uncertainties. 
\label{fig:appfail}}
\end{figure*}

In Figure \ref{fig:appfail}, we show in a color magnitude Hess diagram that the JAGB stars at $1.4<(J-K_s)<2.0$~mag were heavily contaminated by bluer O-rich AGB stars. In this section, we explore several potential explanations for why the photometry and astrometry of the 2MASS sample was too inaccurate to distinguish between the O-rich AGB stars and JAGB stars using the standard color cuts. First, as discussed in Section \ref{subsec:parallaxsigma}, bright ($G<13$~mag) and red stars in the \textit{Gaia} catalog suffer from high parallax errors resulting from their parallax zeropoint offset measurement and uncertainties in the PSF-fitting routines. 
Second, reddening could also be a source of error. In the future, we will explore applying the Wesenheit function  \citep{1982ApJ...253..575M} as a means of correcting the reddening.  Third, \cite{2006AJ....131.1163S} reported large photometric uncertainties on the order of $\sim 0.25$~mag for bright stars with $m_J<5.5$~mag.
Removing stars with $m_J<5.5$~mag or $G<13$~mag was not a viable solution because it would have removed many of the brighter JAGB stars and introduced a systematic bias in $M_J$.

\begin{figure}
\figurenum{A2}
\gridline{\fig{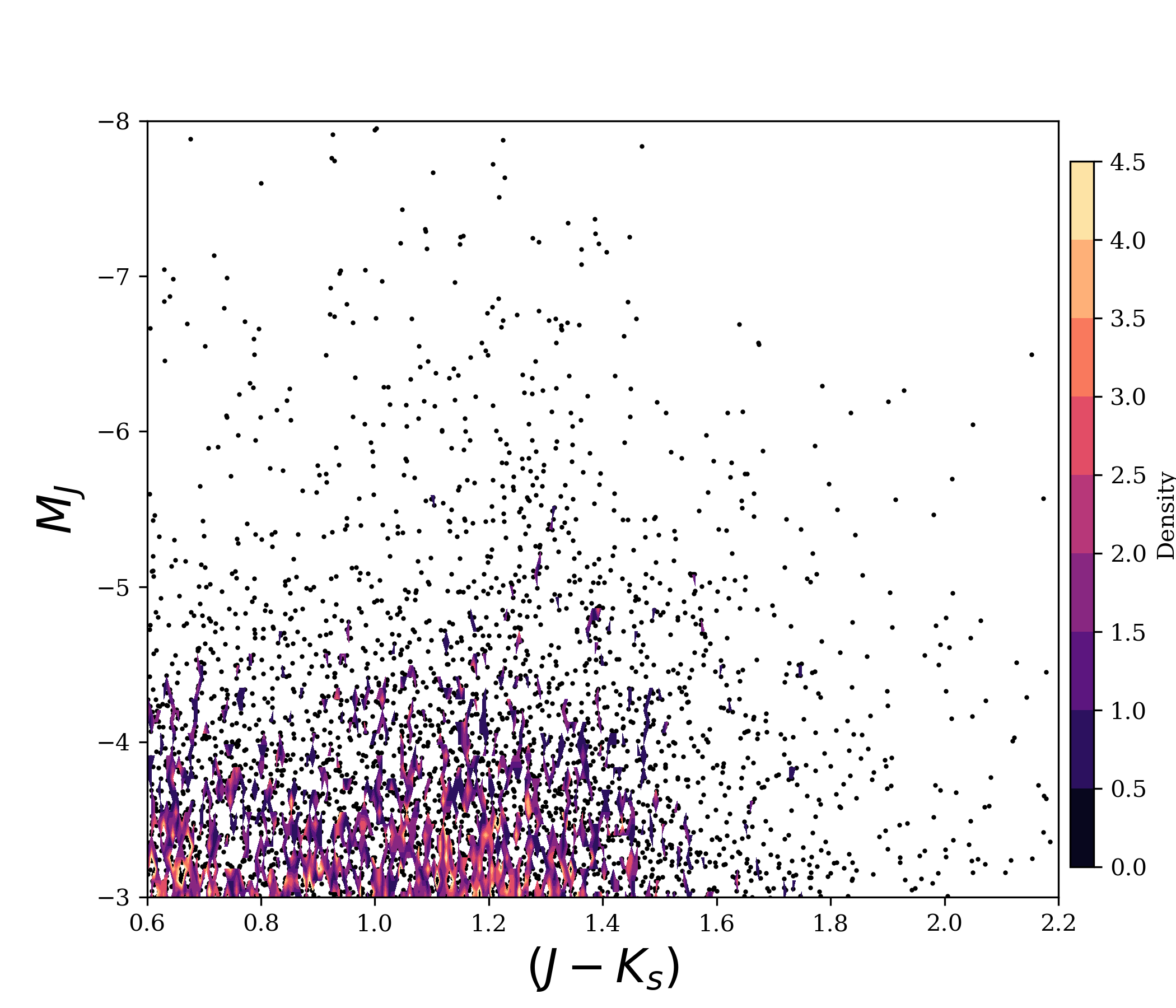}{0.5\textwidth}{}
          \fig{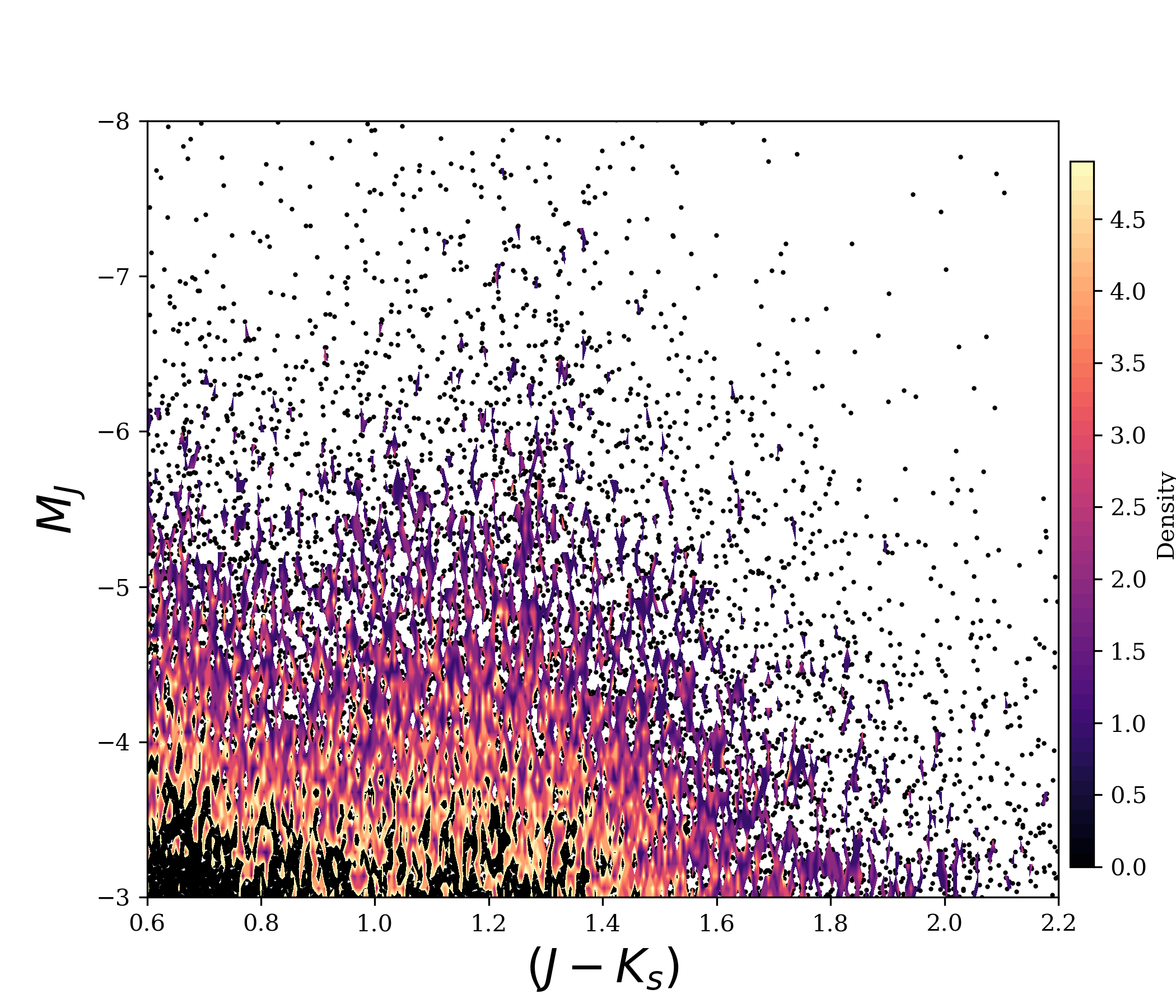}{0.5\textwidth}{}}
\caption{Cleaned $J$ vs. ($J-K_s$) color magnitude Hess diagrams for the 2MASS-selected sample, filtered by distance (left) and fractional parallax uncertainty (right). We determined that attempts to clean the 2MASS photometry using aggressive cuts on distance and fractional parallax uncertainty were still ineffective in isolating the JAGB stars, particularly at the blue end.
\label{fig:photcuts}}
\end{figure}

Attempts to clean our photometry with cuts on distance ($d<1$~kpc) and fractional parallax ($\sigma_{\pi}/\pi<0.05$), shown in Figure  \ref{fig:photcuts} also proved unsuccessful. In both cases, a large fraction of the O-rich AGB stars remained post-cleaning, continuing to make it impossible to distinguish between the O-rich AGB stars and the JAGB stars. Therefore, we found that photometrically selecting a sample of Galactic JAGB stars was not a  viable option at this time,  and for the time being, chose to select JAGB stars from Galactic carbon star catalogs instead.

\end{document}